\documentclass[a4paper]{article}

\usepackage{INTERSPEECH2020}

\usepackage{xcolor}

\usepackage{subcaption}

\usepackage{stfloats}

\usepackage{tikz}
\usetikzlibrary{positioning,decorations.pathreplacing,shapes,arrows,fit}

\pgfdeclarelayer{back}
\pgfsetlayers{back,main}

\makeatletter
\pgfkeys{%
  /tikz/on layer/.code={
    \def\tikz@path@do@at@end{\endpgfonlayer\endgroup\tikz@path@do@at@end}%
    \pgfonlayer{#1}\begingroup%
  }%
}
\makeatother

\tikzset{
  >=latex,
  font=\scriptsize,
  block/.style = {draw},
  model/.style = {draw,dashed,inner sep=3.5mm},
  line/.style = {->,semithick},
  forward/.style = {line},
  backward/.style = {line,densely dotted},
  tcsy/.style = {transform canvas={yshift=#1}},
  tcsx/.style = {transform canvas={xshift=#1}},
}

\title{Pretrained Semantic Speech Embeddings for End-to-End Spoken Language Understanding via Cross-Modal Teacher-Student Learning}
\name{Pavel Denisov, Ngoc Thang Vu}
\address{Institute for Natural Language Processing (IMS), University of Stuttgart, Germany}
\email{\{pavel.denisov, thang.vu\}@ims.uni-stuttgart.de}
\begin{document}

\maketitle
\begin{abstract}
Spoken language understanding is typically based on pipeline architectures including speech recognition and natural language understanding steps. These components are optimized independently to allow usage of available data, but the overall system suffers from error propagation. In this paper, we propose a novel training method that enables pretrained contextual embeddings to process acoustic features. In particular, we extend it with an encoder of pretrained speech recognition systems in order to construct end-to-end spoken language understanding systems. Our proposed method is based on the teacher-student framework across speech and text modalities that aligns the acoustic and the semantic latent spaces. Experimental results in three benchmarks show that our system reaches the performance comparable to the pipeline architecture without using any training data and outperforms it after fine-tuning with ten examples per class on two out of three benchmarks.
\end{abstract}


\section{Introduction}

Recent developments in the fields of electronics, computations and data processing
have led to an increased interest in smart assistants with speech interfaces.
It is likely driven by the fact that usually people can learn to use speech for interaction intuitively
without any special training \cite{pinker1990natural} and make it a primary medium of information exchange.
However, speech poses a major challenge to a machine when it comes to the task of extraction of information intended to be transmitted by a human speaker, also known as Spoken Language Understanding (SLU) \cite{de2008spoken}.
The key difficulty here is that speech is highly variable, e.g. depending on room acoustic, and contains rich information about speakers  \cite{kroger2019privacy}. Some of them are not useful for SLU. The information
extraction task is often performed on the text representation using Natural Language Understanding (NLU) methods \cite{manning1999foundations}, while Automatic Speech Recognition (ASR) systems \cite{baker1975dragon, jelinek1997statistical} convert speech to text.
ASR step removes redundant information from the input and provides some kind of normalized form on the output. At the same time it causes loss of potentially useful information that can not be encoded in the text representation, such as prosody, loudness and speech rate. The operation of finding the most probable sequence of words for speech input is computationally expensive. 
This is partly solved by various heuristics avoiding exploration of less probable hypothesis \cite{viterbi1967error,chorowski2017towards,hori2018end}, what in turn introduces additional errors propagated to NLU component. Finally, the sequential design of pipeline approach leads to unavoidable source of latency, because NLU component can not start its work before ASR is finished, and it is not desirable in the interactive context of smart assistant. The problems of pipeline approach described above can be solved by end-to-end SLU methods.

Existing works on end-to-end SLU modeling either
focus on supervised downstream tasks,
for example dialog act classification \cite{ortega2018lexico},
intent detection \cite{chen2018spoken},
slot filling \cite{tomashenko2019investigating},
independent intent detection and domain classification \cite{serdyuk2018towards} and
joint intent detection, domain classification and slot filling \cite{haghani2018audio},
or target a generic semantic embedding
\cite{chung2018speech2vec,chung2018unsupervised,chuang2019speechbert}
usually inspired by
such successful models as word embeddings Word2Vec \cite{mikolov2013distributed}
and contextual text embeddings BERT \cite{devlin2018bert}.
Highly variable and complex nature of speech leads
to large amounts of both data and computational resources
required for SLU training compared to NLU training,
especially for recently popular approach based on
contextual embeddings.
While data requirements could be satisfied for unsupervised approaches,
computational resources are still a problem.
Fortunately, most of the modern language processing
methods, including ASR and NLU, are based on
neural networks and deep learning.
Deep learning offers an easy way to transfer knowledge
between learned tasks.
This technique is referred as transfer learning
and it is successfully applied in both ASR \cite{vu2013multilingual,kunze2017transfer} and NLU \cite{radford2019language,devlin2018bert}.
Therefore, transfer learning should be a promising
direction to explore for SLU as well.
Several reports
\cite{qian2017exploring,chen2018spoken,haghani2018audio,lugosch2019speech,tomashenko2019investigating,chuang2019speechbert}
indicate that transfer learning from audio modality through
pretraining on ASR task or, alternatively, speech autoencoding,
is helpful for downstream SLU tasks. Transfer
learning from text modality, however, has been
applied only for Speech2Vec \cite{chung2018unsupervised}
and SpeechBERT \cite{chuang2019speechbert} so far.

We propose a novel method that combines parameters transfer
from well trained end-to-end ASR systems \cite{kim2017joint} such as pretrained ESPnet \cite{denisov2019ims} and end-to-end NLU models such as pretrained BERT \cite{devlin2018bert} with
Teacher-Student learning \cite{hinton2015distilling, li2017large} for final alignment of SLU output space to NLU output space
in order to construct end-to-end SLU model
allowing few-shot transfer of downstream tasks
from text to speech.
By doing so, we enable pretrained end-to-end contextual embeddings such as BERT to process acoustic features. 
In particular, we aim to generate fixed length vectors with semantic representation
from speech segments of variable length. Transfer learning
from both text and audio modalities makes our approach
mostly similar to \cite{chuang2019speechbert} and \cite{chung2018unsupervised}.
In this work, we investigate utterance classification task
and focus on zero-shot and few-shot cases,
but the described method could be
adopted to many types of SLU tasks.
Although previous works described a number of experiments
for such utterance classification tasks
as dialog act classification \cite{ortega2019context}
and intent classification \cite{lugosch2019speech},
and we use the same datasets for the evaluation,
we do not compare our results directly to these works,
as this is outside of the scope of our work.

\section{Method}

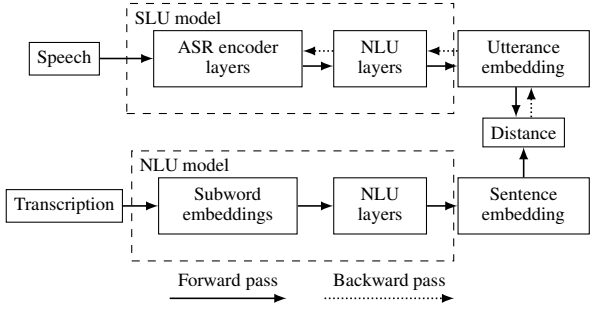
\begin{figure}[!t]
\centering
\begin{tikzpicture}

\matrix[column sep=4mm, row sep=4mm] (m) {
\node[block] (speech) {Speech}; &
\node[block] (asr-encoder) {
\begin{tabular}{c}
ASR encoder \\
layers \\
\end{tabular}
}; &
\node[block] (nlu-slu) {
\begin{tabular}{c}
NLU \\
layers \\
\end{tabular}
}; &
\node[block] (utterance-embedding) {
\begin{tabular}{c}
Utterance \\
embedding \\
\end{tabular}
}; \\
& & & \node[block] (distance) {Distance}; \\
\node[block] (transcription) {Transcription}; &
\node[block] (subword-embeddings) {
\begin{tabular}{c}
Subword \\
embeddings \\
\end{tabular}
}; &
\node[block] (nlu-nlu) {
\begin{tabular}{c}
NLU \\
layers \\
\end{tabular}
}; &
\node[block] (sentence-embedding) {
\begin{tabular}{c}
Sentence \\
embedding \\
\end{tabular}
}; \\
};

\begin{pgfonlayer}{back}
\node[fit=(asr-encoder) (nlu-slu), model] (slu-model) {};
\node[below right] at (slu-model.north west) {SLU model};
\node[fit=(subword-embeddings) (nlu-nlu), model] (nlu-model) {};
\node[below right] at (nlu-model.north west) {NLU model};
\end{pgfonlayer}

\draw[forward] (speech) to (asr-encoder);
\draw[forward] (asr-encoder) edge [tcsy=-1mm] (nlu-slu);
\draw[forward] (nlu-slu) edge [tcsy=-1mm] (utterance-embedding);
\draw[forward] (utterance-embedding) edge [tcsx=-1mm] (distance);

\draw[backward] (nlu-slu) edge [tcsy=1mm] (asr-encoder);
\draw[backward] (utterance-embedding) edge [tcsy=1mm] (nlu-slu);
\draw[backward] (distance) edge [tcsx=1mm] (utterance-embedding);

\draw[forward] (transcription) to (subword-embeddings);
\draw[forward] (subword-embeddings) to (nlu-nlu);
\draw[forward] (nlu-nlu) to (sentence-embedding);
\draw[forward] (sentence-embedding) to (distance);

\node[below = 4mm of subword-embeddings] (forward) {Forward pass};
\node[right = 5mm of forward] (backward) {Backward pass};

\draw[forward] (forward.south west) to (forward.south east);
\draw[backward] (backward.south west) to (backward.south east);

\end{tikzpicture}
\caption{End-to-end SLU using cross-modal T-S learning.}
\label{fig:method}
\end{figure}

Figure \ref{fig:method} provides the overview of the proposed method.
Our SLU model is a combination of two pretrained models. First, we
use Encoder block of pretrained end-to-end ASR model \cite{kim2017joint} in order to covert acoustic features of speech signal to hidden representation.
Second, we feed the hidden representation
through a learnable linear mapping
to pretrained masked language model \cite{devlin2018bert},
fine-tuned to produce semantic sentence embedding,
which serves as NLU model.
Finally, we utilize teacher-student learning method in order to align
output of our SLU model to output of pretrained NLU model.
Both ASR and NLU models are based on Transformer architecture \cite{vaswani2017attention}
widely used for sequence processing.

\subsection{End-to-end ASR}

ASR model implements sequence-to-sequence
approach and contains two major blocks,
encoder and decoder. Encoder takes sequence
$X$ with acoustic features and outputs encoded sequence
$X_\epsilon$ with hidden representation.
Decoder takes the encoded sequence $X_\epsilon$ on input
and outputs target sequence $Y$ with text tokens
representing transcription of the input utterance.
ASR model is trained to minimize weighted sum
of cross-entropy objective function calculated from
decoder output $\hat{Y}$ and ground truth transcription $Y$
with CTC objective function calculated from
learnable linear mapping of encoder output $X_\epsilon$
and ground truth transcription $Y$.

\subsection{NLU}

NLU model is a neural network that takes sequence $X$
with text tokens on input and produces encoded sequence $X_\epsilon$.
Semantic sentence embedding vector $y$ is obtained
by applying pooling operation to the encoded sequence $X_\epsilon$.
The model's parameters are initially pretrained with the tasks of
masked token and next sentence prediction
from the encoded sequence $X_\epsilon$ representing
contextual text token embeddings, as it is done
with BERT model \cite{devlin2018bert}.
After that, the model is extended with pooling operation
over the encoded sequence $X_\epsilon$, producing pooled output $y$,
and is fine-tuned on specialized datasets to encode more
semantic information to the pooled output $y$.

\subsection{Teacher-Student learning}

Teacher-Student learning minimizes distance-based
objective function between outputs of two models on
same or equivalent inputs with the aim
to update Student model's parameters so
that its output becomes more similar to the
output of Teacher model. The parameters
of Teacher model are not updated during this process.
The final stage of our method is the alignment of
SLU output to NLU output with Teacher-Student learning method,
where SLU model consumes speech recordings and plays the Student role,
while NLU model consumes ground truth transcriptions
and plays the Teacher role.

\section{Experimental setup}

\subsection{ASR model}

We adopt the latest LibriSpeech
recipe \cite{karita2019comparative} from ESPnet toolkit.
Transformer network has attention dimension 512,
feed-forward inner dimension 2048, 8 heads,
12 blocks in the encoder and 6 blocks in the decoder.
Input features are 80-dimensional log Mel filterbank coefficients
with 3-dimensional pitch value, frame size is 25 ms and shift is 10 ms.
Output labels are 100 subword units, automatically
learned with unigram language model algorithm \cite{kudo2018subword} from
lowercased concatenation of LibriSpeech and TED-LIUM LM training data with
transcriptions of the acoustic training data.
Training data combines LibriSpeech, Switchboard,
TED-LIUM~3, AMI, WSJ, Common~Voice~3, SWC, VoxForge and \mbox{M-AILABS}
datasets with a total amount of 3249 hours.
Validation data combines validation subsets of LibriSpeech, TED-LIUM~3
and AMI datasets with a total amount of 38 hours.
The training is performed on 4 GPUs using Adam optimizer
and square root learning rate scheduling \cite{vaswani2017attention}
with 25,000 warmup steps and learning rate coefficient 10.
SpecAugment data augmentation method \cite{park2019specaugment} is applied dynamically
during each batch generation.
The model is trained for 24 epochs and evaluated
on the validation data after each epoch.
The final model is obtained by averaging the parameters
of the seven best performing models.

\subsection{NLU model}

We use pretrained \texttt{bert-base-nli-stsb-mean-tokens}
Sentence-BERT model \cite{reimers2019sentence}.
The model itself is fine-tuned from the well-known pretrained
\texttt{bert-base-uncased} model \cite{devlin2018bert}.
Transformer network has attention dimension 768,
feed-forward inner dimension 3072, 12 heads and 12 blocks.
Input text is tokenized to 30,000 subword units.
The model is pretrained with masked LM and next sentence prediction tasks
on BooksCorpus and English Wikipedia datasets.
Pooling operation \texttt{MEAN} is added to obtain the sentence embedding $y$
from the encoded sequence $X_\epsilon$.
The sentence embedding is first fine-tuned on SNLI and MultiNLI datasets for 3-way
classification between \emph{contradiction},
\emph{entailment} and \emph{neutral} classes
for a given pair of sentences using cross-entropy objective function.
After that, the sentence embedding is fine-tuned on STSb dataset
for prediction of cosine similarity for a given pair of sentences
using mean-squared-error objective function.

\subsection{SLU model}

SLU model is constructed by combining ASR model's encoder with self-attention
blocks of NLU model, so that NLU model receives the hidden representation
from ASR encoder instead of the output of input embedding layer of NLU model.
Linear layer is added between ASR encoder and NLU blocks to map the dimension
of hidden representation from 512 to 768. Fine-tuning is performed using
Teacher-Student approach by minimizing
the distance between output of SLU model for speech recordings
and output of NLU model for corresponding transcriptions.
We conduct fine-tuning experiments with cosine,
L2 and L1 distance based objective functions.
SLU model acts as a Student, and we select empirically,
which parameters to update during
the fine-tuning. NLU model acts as a Teacher, and we freeze its parameters.
We employ smaller acoustic dataset consisting of LibriTTS, Common~Voice~3,
and \mbox{M-AILABS} corpora with a total amount of 1453 hours for the fine-tuning.
Our motivation here is to utilize richer transcriptions with punctuation
available in these datasets and to supply NLU model with extra information
for potentially semantically finer sentence embeddings.
We use the transcriptions as is and do not apply any text preprocessing
that is usually done in ASR training, including our end-to-end ASR model.
Validation data is the validation subset of LibriTTS corpus
with a total duration of 15 hours.
We do not apply SpecAugment during the fine-tuning, because
it yielded worse results in our early experiments.
 
\subsection{Evaluation}

SLU model is evaluated on two downstream tasks,
dialog act (DA) classification and intent classification,
both of which are utterance classification tasks.
DA classification is evaluated on two corpora:
ICSI Meeting Recorder Dialog Act Corpus (MRDA)
and NXT-format Switchboard Corpus (SwDA).
Intent classification is evaluated on Fluent
Speech Commands (FSC) corpus. Table \ref{tab:datasets} summarizes the datasets.

\begin{table}[h]
  \caption{ SLU evaluation datasets }
  \label{tab:datasets}
  \centering
  \footnotesize
  \begin{tabular*}{0.82\columnwidth}{l|c|c|c|c}
    \noalign{\hrule height 1pt}
    Dataset & Number of & \multicolumn{3}{|c}{Number of utterances} \\
    \cline{3-5}
    & classes & Train & Valid & Test \\
    \hline
    SwBD & 42 & 97,756 & 8,591 & 2,507 \\
    MRDA & 6 & 77,596 & 15,721 & 15,398 \\
    FSC & 31 & 23,132 & 3,118 & 3,793 \\
    \noalign{\hrule height 1pt}
  \end{tabular*}
\end{table}

In order to perform utterance classification,
we first train a one layer feed-forward classifier
on sentence embeddings, produced by the NLU model
from the ground truth transcriptions of training subset,
using cross-entropy objective function. After that,
we test the classifier on semantic utterance embeddings,
extracted from the recordings of testing subset using the SLU model.
We report accuracy values
as a percentage of correctly classified utterances from
the total number of utterances.

\subsection{Baseline}
Traditional approach to SLU tasks is a pipeline of
ASR followed by NLU, and we adopt it as a baseline
while employing the same ASR and NLU models as
in the rest of the experiments.
Table \ref{tab:baseline} reports the results
of NLU on ASR output as well as on the ground truth
transcriptions. The ground truth results
represent an upper bound of accuracy
achievable on these datasets with NLU model we use
in case of perfect transcriptions on ASR output.
The effect of imperfect ASR output varies between
datasets depending on the difficulty of recording conditions,
the differences between formats of manual transcriptions
used to train the classifiers
and the tolerance of the downstream tasks
to the type of noise that ASR introduces. Amount of errors in ASR output is  indicated by Word Error Rate (WER), which is also reported in the table. We select the best performing hyperparameters for the classifier training, but do not fine-tune NLU component for the downstream tasks, because our main goal is SLU as generic speech equivalent for NLU rather then the best possible model for some particular downstream task.

\begin{table}[h]
  \caption{ Accuracy of NLU on ASR output and on the ground truth transcriptions and WER of ASR}
  \label{tab:baseline}
  \centering
  \footnotesize
  \begin{tabular*}{0.69\columnwidth}{l|c|c|c}
    \noalign{\hrule height 1pt}
    Transcriptions & \multicolumn{3}{|c}{Accuracy on Test, \%} \\
    \cline{2-4}
    & SwBD & MRDA & FSC \\
    \hline
    Ground truth & 71.72 & 77.72 & 100.0 \\
    ASR output & 57.23 & 64.06 & 94.57 \\
    \hline
     & \multicolumn{3}{|c}{WER on Test, \%} \\
     \hline
    ASR output   & 28.0 & 29.7 & 7.9 \\
    \noalign{\hrule height 1pt}
  \end{tabular*}
\end{table}

\section{Results}

\subsection{Initial fine-tuning by Teacher-Student learning}

\subsubsection{Layers for fine-tuning}

Our first set of experiments is designed to determine which
layers of SLU model should be fine-tuned after the combination
of parameters transferred from ASR encoder and NLU.
As mentioned before, we insert a linear mapping layer between
former ASR and NLU layers because of the difference in dimensionality.
It is initialized randomly and its parameters are always updated
during the fine-tuning step. In addition to that, we try
to fine-tune various amount of layers closest to the mapping
layer, meaning top layers of former ASR encoder
and bottom layers of former NLU.
We do so, because for these layers
the output (for ASR encoder) or the input (for NLU)
is expected to change after the parameters transfer
in contrast to bottom layers of former ASR encoder
and top layers of former NLU, where input and output
should not change.

We run fine-tuning for 10 epochs using
square root learning rate scheduling \cite{vaswani2017attention}
with 300,000 warmup steps and learning rate coefficient 50,
and use cosine distance based objective function.
The results are given in Table~\ref{tab:layers}.
While it is not completely clear how many layers
should be fine-tuned, we can conclusively tell
that fine-tuning of former ASR encoder layers
is more beneficial than former NLU layers.
We decide to fine-tune the two top former ASR encoder layers.
The results also illustrate that the optimization of SLU model
for smaller distance of its output
from the output of NLU model
is general enough and translates to accuracy improvements
in the downstream tasks, although not in all cases.

\begin{table}[h]
  \caption{Effect of layers fine-tuning}
  \label{tab:layers}
  \centering
  \footnotesize
  \begin{tabular*}{0.91\columnwidth}{l|c|c|c|c|c}
    \noalign{\hrule height 1pt}
	  ASR & NLU & \multicolumn{3}{|c|}{Accuracy on Test, \%} & Validation \\
    \cline{3-5}
	  layers &  layers & SwBD & MRDA & FSC & loss \\
    \hline
	  0 & 0 & 43.76 & 56.08 & 68.07 & 0.26 \\
	  0 & 1 & 37.61 & 56.47 & 85.53 & 0.19 \\
	  1 & 0 & 52.37 & \textbf{60.21} & 86.42 & 0.16 \\
	  1 & 1 & 52.05 & 58.32 & \textbf{86.82} & 0.17 \\
	  2 & 0 & 52.93 & 59.42 & 85.76 & \textbf{0.15} \\
	  3 & 0 & \textbf{53.81} & 58.90 & 85.53 & 0.16 \\
    \noalign{\hrule height 1pt}
  \end{tabular*}
\end{table}

\subsubsection{Learning rate schedule}

After deciding which layers to fine-tune,
we run a series of experiments to determine the
best learning rate schedule. Table \ref{tab:lr} presents
the combinations of learning rate constant
and number of warmup steps explored by us.
When we increase number of warmup steps,
we notice positive effect from slower
learning rate ramp up.
However, as number number of warmup steps
becomes close to the total number of fine-tuning
steps, we have to increase number of epochs from 10 to 20
in order to see the whole fine-tuning process.

\begin{table}[h]
  \caption{Effect of learning rate schedule}
  \label{tab:lr}
  \centering
  \footnotesize
  \setlength{\tabcolsep}{4pt}
  \begin{tabular*}{1.02\columnwidth}{l|c|c|c|c|c|c}
    \noalign{\hrule height 1pt}
	  Warmup & LR & Epochs & \multicolumn{3}{|c|}{Accuracy on Test, \%} & Validation \\
    \cline{4-6}
	  steps &  constant &  & SwBD & MRDA & FSC & loss \\
    \hline
     300,000 & 50 & 10 & 52.93 & 59.42 & 85.76 & 0.15 \\
	  600,000 & 50 & 10 & 51.18 & 59.95 & 86.84 & 0.14 \\
	  600,000 & 50 & 20 & 54.00 & \textbf{60.12} & 88.64 & 0.14 \\
	  700,000 & 50 & 20 & 51.89 & 58.37 & 88.24 & 0.14 \\
	  700,000 & 70 & 20 & 53.73 & 59.67 & 88.08 & 0.14 \\
	  700,000 & 30 & 20 & \textbf{55.56} & 59.64 & \textbf{89.45} & \textbf{0.13} \\
    \noalign{\hrule height 1pt}
  \end{tabular*}
\end{table}

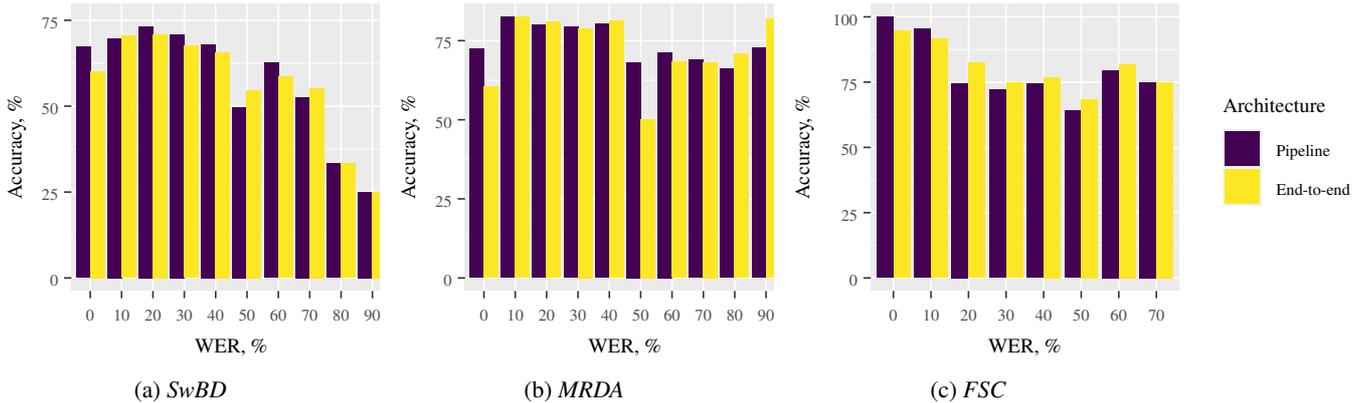
\begin{figure*}[ht]
	\hspace*{-0.3in}
	\begin{subfigure}{.3\textwidth}
	\centering
\begin{tikzpicture}[x=1pt,y=1pt]
\definecolor{fillColor}{RGB}{255,255,255}
\path[use as bounding box,fill=fillColor,fill opacity=0.00] (0,0) rectangle (216.81,144.54);
\begin{scope}
\path[clip] (  0.00,  0.00) rectangle (216.81,144.54);
\definecolor{drawColor}{RGB}{255,255,255}
\definecolor{fillColor}{RGB}{255,255,255}

\path[draw=drawColor,line width= 0.6pt,line join=round,line cap=round,fill=fillColor] (  0.00,  0.00) rectangle (216.81,144.54);
\end{scope}
\begin{scope}
\path[clip] ( 31.71, 30.69) rectangle (151.39,139.04);
\definecolor{fillColor}{gray}{0.92}

\path[fill=fillColor] ( 31.71, 30.69) rectangle (151.39,139.04);
\definecolor{drawColor}{RGB}{255,255,255}

\path[draw=drawColor,line width= 0.3pt,line join=round] ( 31.71, 51.81) --
	(151.39, 51.81);

\path[draw=drawColor,line width= 0.3pt,line join=round] ( 31.71, 84.21) --
	(151.39, 84.21);

\path[draw=drawColor,line width= 0.3pt,line join=round] ( 31.71,116.62) --
	(151.39,116.62);

\path[draw=drawColor,line width= 0.6pt,line join=round] ( 31.71, 35.61) --
	(151.39, 35.61);

\path[draw=drawColor,line width= 0.6pt,line join=round] ( 31.71, 68.01) --
	(151.39, 68.01);

\path[draw=drawColor,line width= 0.6pt,line join=round] ( 31.71,100.42) --
	(151.39,100.42);

\path[draw=drawColor,line width= 0.6pt,line join=round] ( 31.71,132.82) --
	(151.39,132.82);

\path[draw=drawColor,line width= 0.6pt,line join=round] ( 38.75, 30.69) --
	( 38.75,139.04);

\path[draw=drawColor,line width= 0.6pt,line join=round] ( 50.49, 30.69) --
	( 50.49,139.04);

\path[draw=drawColor,line width= 0.6pt,line join=round] ( 62.22, 30.69) --
	( 62.22,139.04);

\path[draw=drawColor,line width= 0.6pt,line join=round] ( 73.95, 30.69) --
	( 73.95,139.04);

\path[draw=drawColor,line width= 0.6pt,line join=round] ( 85.69, 30.69) --
	( 85.69,139.04);

\path[draw=drawColor,line width= 0.6pt,line join=round] ( 97.42, 30.69) --
	( 97.42,139.04);

\path[draw=drawColor,line width= 0.6pt,line join=round] (109.15, 30.69) --
	(109.15,139.04);

\path[draw=drawColor,line width= 0.6pt,line join=round] (120.89, 30.69) --
	(120.89,139.04);

\path[draw=drawColor,line width= 0.6pt,line join=round] (132.62, 30.69) --
	(132.62,139.04);

\path[draw=drawColor,line width= 0.6pt,line join=round] (144.35, 30.69) --
	(144.35,139.04);
\definecolor{fillColor}{RGB}{68,1,84}

\path[fill=fillColor] ( 33.47, 35.61) rectangle ( 38.75,122.75);
\definecolor{fillColor}{RGB}{253,231,37}

\path[fill=fillColor] ( 38.75, 35.61) rectangle ( 44.03,113.60);
\definecolor{fillColor}{RGB}{68,1,84}

\path[fill=fillColor] ( 45.21, 35.61) rectangle ( 50.49,125.99);
\definecolor{fillColor}{RGB}{253,231,37}

\path[fill=fillColor] ( 50.49, 35.61) rectangle ( 55.77,126.98);
\definecolor{fillColor}{RGB}{68,1,84}

\path[fill=fillColor] ( 56.94, 35.61) rectangle ( 62.22,130.46);
\definecolor{fillColor}{RGB}{253,231,37}

\path[fill=fillColor] ( 62.22, 35.61) rectangle ( 67.50,127.45);
\definecolor{fillColor}{RGB}{68,1,84}

\path[fill=fillColor] ( 68.67, 35.61) rectangle ( 73.95,127.57);
\definecolor{fillColor}{RGB}{253,231,37}

\path[fill=fillColor] ( 73.95, 35.61) rectangle ( 79.23,123.22);
\definecolor{fillColor}{RGB}{68,1,84}

\path[fill=fillColor] ( 80.41, 35.61) rectangle ( 85.69,123.71);
\definecolor{fillColor}{RGB}{253,231,37}

\path[fill=fillColor] ( 85.69, 35.61) rectangle ( 90.97,120.67);
\definecolor{fillColor}{RGB}{68,1,84}

\path[fill=fillColor] ( 92.14, 35.61) rectangle ( 97.42,100.01);
\definecolor{fillColor}{RGB}{253,231,37}

\path[fill=fillColor] ( 97.42, 35.61) rectangle (102.70,106.45);
\definecolor{fillColor}{RGB}{68,1,84}

\path[fill=fillColor] (103.87, 35.61) rectangle (109.15,116.78);
\definecolor{fillColor}{RGB}{253,231,37}

\path[fill=fillColor] (109.15, 35.61) rectangle (114.43,111.54);
\definecolor{fillColor}{RGB}{68,1,84}

\path[fill=fillColor] (115.61, 35.61) rectangle (120.89,103.83);
\definecolor{fillColor}{RGB}{253,231,37}

\path[fill=fillColor] (120.89, 35.61) rectangle (126.17,107.24);
\definecolor{fillColor}{RGB}{68,1,84}

\path[fill=fillColor] (127.34, 35.61) rectangle (132.62, 78.81);
\definecolor{fillColor}{RGB}{253,231,37}

\path[fill=fillColor] (132.62, 35.61) rectangle (137.90, 78.81);
\definecolor{fillColor}{RGB}{68,1,84}

\path[fill=fillColor] (139.07, 35.61) rectangle (144.35, 68.01);
\definecolor{fillColor}{RGB}{253,231,37}

\path[fill=fillColor] (144.35, 35.61) rectangle (149.63, 68.01);
\end{scope}
\begin{scope}
\path[clip] (  0.00,  0.00) rectangle (216.81,144.54);
\definecolor{drawColor}{gray}{0.30}

\node[text=drawColor,anchor=base east,inner sep=0pt, outer sep=0pt, scale=  0.88] at ( 26.76, 32.58) {0};

\node[text=drawColor,anchor=base east,inner sep=0pt, outer sep=0pt, scale=  0.88] at ( 26.76, 64.98) {25};

\node[text=drawColor,anchor=base east,inner sep=0pt, outer sep=0pt, scale=  0.88] at ( 26.76, 97.39) {50};

\node[text=drawColor,anchor=base east,inner sep=0pt, outer sep=0pt, scale=  0.88] at ( 26.76,129.79) {75};
\end{scope}
\begin{scope}
\path[clip] (  0.00,  0.00) rectangle (216.81,144.54);
\definecolor{drawColor}{gray}{0.20}

\path[draw=drawColor,line width= 0.6pt,line join=round] ( 28.96, 35.61) --
	( 31.71, 35.61);

\path[draw=drawColor,line width= 0.6pt,line join=round] ( 28.96, 68.01) --
	( 31.71, 68.01);

\path[draw=drawColor,line width= 0.6pt,line join=round] ( 28.96,100.42) --
	( 31.71,100.42);

\path[draw=drawColor,line width= 0.6pt,line join=round] ( 28.96,132.82) --
	( 31.71,132.82);
\end{scope}
\begin{scope}
\path[clip] (  0.00,  0.00) rectangle (216.81,144.54);
\definecolor{drawColor}{gray}{0.20}

\path[draw=drawColor,line width= 0.6pt,line join=round] ( 38.75, 27.94) --
	( 38.75, 30.69);

\path[draw=drawColor,line width= 0.6pt,line join=round] ( 50.49, 27.94) --
	( 50.49, 30.69);

\path[draw=drawColor,line width= 0.6pt,line join=round] ( 62.22, 27.94) --
	( 62.22, 30.69);

\path[draw=drawColor,line width= 0.6pt,line join=round] ( 73.95, 27.94) --
	( 73.95, 30.69);

\path[draw=drawColor,line width= 0.6pt,line join=round] ( 85.69, 27.94) --
	( 85.69, 30.69);

\path[draw=drawColor,line width= 0.6pt,line join=round] ( 97.42, 27.94) --
	( 97.42, 30.69);

\path[draw=drawColor,line width= 0.6pt,line join=round] (109.15, 27.94) --
	(109.15, 30.69);

\path[draw=drawColor,line width= 0.6pt,line join=round] (120.89, 27.94) --
	(120.89, 30.69);

\path[draw=drawColor,line width= 0.6pt,line join=round] (132.62, 27.94) --
	(132.62, 30.69);

\path[draw=drawColor,line width= 0.6pt,line join=round] (144.35, 27.94) --
	(144.35, 30.69);
\end{scope}
\begin{scope}
\path[clip] (  0.00,  0.00) rectangle (216.81,144.54);
\definecolor{drawColor}{gray}{0.30}

\node[text=drawColor,anchor=base,inner sep=0pt, outer sep=0pt, scale=  0.88] at ( 38.75, 19.68) {0};

\node[text=drawColor,anchor=base,inner sep=0pt, outer sep=0pt, scale=  0.88] at ( 50.49, 19.68) {10};

\node[text=drawColor,anchor=base,inner sep=0pt, outer sep=0pt, scale=  0.88] at ( 62.22, 19.68) {20};

\node[text=drawColor,anchor=base,inner sep=0pt, outer sep=0pt, scale=  0.88] at ( 73.95, 19.68) {30};

\node[text=drawColor,anchor=base,inner sep=0pt, outer sep=0pt, scale=  0.88] at ( 85.69, 19.68) {40};

\node[text=drawColor,anchor=base,inner sep=0pt, outer sep=0pt, scale=  0.88] at ( 97.42, 19.68) {50};

\node[text=drawColor,anchor=base,inner sep=0pt, outer sep=0pt, scale=  0.88] at (109.15, 19.68) {60};

\node[text=drawColor,anchor=base,inner sep=0pt, outer sep=0pt, scale=  0.88] at (120.89, 19.68) {70};

\node[text=drawColor,anchor=base,inner sep=0pt, outer sep=0pt, scale=  0.88] at (132.62, 19.68) {80};

\node[text=drawColor,anchor=base,inner sep=0pt, outer sep=0pt, scale=  0.88] at (144.35, 19.68) {90};
\end{scope}
\begin{scope}
\path[clip] (  0.00,  0.00) rectangle (216.81,144.54);
\definecolor{drawColor}{RGB}{0,0,0}

\node[text=drawColor,anchor=base,inner sep=0pt, outer sep=0pt, scale=  1.10] at ( 91.55,  7.64) {WER, \%};
\end{scope}
\begin{scope}
\path[clip] (  0.00,  0.00) rectangle (216.81,144.54);
\definecolor{drawColor}{RGB}{0,0,0}

\node[text=drawColor,rotate= 90.00,anchor=base,inner sep=0pt, outer sep=0pt, scale=  1.10] at ( 13.08, 84.86) {Accuracy, \%};
\end{scope}
\begin{scope}
\path[clip] (  0.00,  0.00) rectangle (216.81,144.54);
\definecolor{fillColor}{RGB}{255,255,255}

\path[fill=fillColor] (162.39, 57.30) rectangle (211.31,112.42);
\end{scope}
\end{tikzpicture}
	\vspace*{-0.2in}
	\caption{SwBD}
	\end{subfigure}
	\begin{subfigure}{.3\textwidth}
	\centering
\begin{tikzpicture}[x=1pt,y=1pt]
\definecolor{fillColor}{RGB}{255,255,255}
\path[use as bounding box,fill=fillColor,fill opacity=0.00] (0,0) rectangle (216.81,144.54);
\begin{scope}
\path[clip] (  0.00,  0.00) rectangle (216.81,144.54);
\definecolor{drawColor}{RGB}{255,255,255}
\definecolor{fillColor}{RGB}{255,255,255}

\path[draw=drawColor,line width= 0.6pt,line join=round,line cap=round,fill=fillColor] (  0.00,  0.00) rectangle (216.81,144.54);
\end{scope}
\begin{scope}
\path[clip] ( 31.71, 30.69) rectangle (151.39,139.04);
\definecolor{fillColor}{gray}{0.92}

\path[fill=fillColor] ( 31.71, 30.69) rectangle (151.39,139.04);
\definecolor{drawColor}{RGB}{255,255,255}

\path[draw=drawColor,line width= 0.3pt,line join=round] ( 31.71, 50.53) --
	(151.39, 50.53);

\path[draw=drawColor,line width= 0.3pt,line join=round] ( 31.71, 80.37) --
	(151.39, 80.37);

\path[draw=drawColor,line width= 0.3pt,line join=round] ( 31.71,110.21) --
	(151.39,110.21);

\path[draw=drawColor,line width= 0.6pt,line join=round] ( 31.71, 35.61) --
	(151.39, 35.61);

\path[draw=drawColor,line width= 0.6pt,line join=round] ( 31.71, 65.45) --
	(151.39, 65.45);

\path[draw=drawColor,line width= 0.6pt,line join=round] ( 31.71, 95.29) --
	(151.39, 95.29);

\path[draw=drawColor,line width= 0.6pt,line join=round] ( 31.71,125.13) --
	(151.39,125.13);

\path[draw=drawColor,line width= 0.6pt,line join=round] ( 38.75, 30.69) --
	( 38.75,139.04);

\path[draw=drawColor,line width= 0.6pt,line join=round] ( 50.49, 30.69) --
	( 50.49,139.04);

\path[draw=drawColor,line width= 0.6pt,line join=round] ( 62.22, 30.69) --
	( 62.22,139.04);

\path[draw=drawColor,line width= 0.6pt,line join=round] ( 73.95, 30.69) --
	( 73.95,139.04);

\path[draw=drawColor,line width= 0.6pt,line join=round] ( 85.69, 30.69) --
	( 85.69,139.04);

\path[draw=drawColor,line width= 0.6pt,line join=round] ( 97.42, 30.69) --
	( 97.42,139.04);

\path[draw=drawColor,line width= 0.6pt,line join=round] (109.15, 30.69) --
	(109.15,139.04);

\path[draw=drawColor,line width= 0.6pt,line join=round] (120.89, 30.69) --
	(120.89,139.04);

\path[draw=drawColor,line width= 0.6pt,line join=round] (132.62, 30.69) --
	(132.62,139.04);

\path[draw=drawColor,line width= 0.6pt,line join=round] (144.35, 30.69) --
	(144.35,139.04);
\definecolor{fillColor}{RGB}{68,1,84}

\path[fill=fillColor] ( 33.47, 35.61) rectangle ( 38.75,121.98);
\definecolor{fillColor}{RGB}{253,231,37}

\path[fill=fillColor] ( 38.75, 35.61) rectangle ( 44.03,107.75);
\definecolor{fillColor}{RGB}{68,1,84}

\path[fill=fillColor] ( 45.21, 35.61) rectangle ( 50.49,134.03);
\definecolor{fillColor}{RGB}{253,231,37}

\path[fill=fillColor] ( 50.49, 35.61) rectangle ( 55.77,134.11);
\definecolor{fillColor}{RGB}{68,1,84}

\path[fill=fillColor] ( 56.94, 35.61) rectangle ( 62.22,131.01);
\definecolor{fillColor}{RGB}{253,231,37}

\path[fill=fillColor] ( 62.22, 35.61) rectangle ( 67.50,132.14);
\definecolor{fillColor}{RGB}{68,1,84}

\path[fill=fillColor] ( 68.67, 35.61) rectangle ( 73.95,130.43);
\definecolor{fillColor}{RGB}{253,231,37}

\path[fill=fillColor] ( 73.95, 35.61) rectangle ( 79.23,129.80);
\definecolor{fillColor}{RGB}{68,1,84}

\path[fill=fillColor] ( 80.41, 35.61) rectangle ( 85.69,131.38);
\definecolor{fillColor}{RGB}{253,231,37}

\path[fill=fillColor] ( 85.69, 35.61) rectangle ( 90.97,132.62);
\definecolor{fillColor}{RGB}{68,1,84}

\path[fill=fillColor] ( 92.14, 35.61) rectangle ( 97.42,117.02);
\definecolor{fillColor}{RGB}{253,231,37}

\path[fill=fillColor] ( 97.42, 35.61) rectangle (102.70, 95.51);
\definecolor{fillColor}{RGB}{68,1,84}

\path[fill=fillColor] (103.87, 35.61) rectangle (109.15,120.74);
\definecolor{fillColor}{RGB}{253,231,37}

\path[fill=fillColor] (109.15, 35.61) rectangle (114.43,117.26);
\definecolor{fillColor}{RGB}{68,1,84}

\path[fill=fillColor] (115.61, 35.61) rectangle (120.89,117.98);
\definecolor{fillColor}{RGB}{253,231,37}

\path[fill=fillColor] (120.89, 35.61) rectangle (126.17,117.00);
\definecolor{fillColor}{RGB}{68,1,84}

\path[fill=fillColor] (127.34, 35.61) rectangle (132.62,114.48);
\definecolor{fillColor}{RGB}{253,231,37}

\path[fill=fillColor] (132.62, 35.61) rectangle (137.90,120.16);
\definecolor{fillColor}{RGB}{68,1,84}

\path[fill=fillColor] (139.07, 35.61) rectangle (144.35,122.42);
\definecolor{fillColor}{RGB}{253,231,37}

\path[fill=fillColor] (144.35, 35.61) rectangle (149.63,133.27);
\end{scope}
\begin{scope}
\path[clip] (  0.00,  0.00) rectangle (216.81,144.54);
\definecolor{drawColor}{gray}{0.30}

\node[text=drawColor,anchor=base east,inner sep=0pt, outer sep=0pt, scale=  0.88] at ( 26.76, 32.58) {0};

\node[text=drawColor,anchor=base east,inner sep=0pt, outer sep=0pt, scale=  0.88] at ( 26.76, 62.42) {25};

\node[text=drawColor,anchor=base east,inner sep=0pt, outer sep=0pt, scale=  0.88] at ( 26.76, 92.26) {50};

\node[text=drawColor,anchor=base east,inner sep=0pt, outer sep=0pt, scale=  0.88] at ( 26.76,122.10) {75};
\end{scope}
\begin{scope}
\path[clip] (  0.00,  0.00) rectangle (216.81,144.54);
\definecolor{drawColor}{gray}{0.20}

\path[draw=drawColor,line width= 0.6pt,line join=round] ( 28.96, 35.61) --
	( 31.71, 35.61);

\path[draw=drawColor,line width= 0.6pt,line join=round] ( 28.96, 65.45) --
	( 31.71, 65.45);

\path[draw=drawColor,line width= 0.6pt,line join=round] ( 28.96, 95.29) --
	( 31.71, 95.29);

\path[draw=drawColor,line width= 0.6pt,line join=round] ( 28.96,125.13) --
	( 31.71,125.13);
\end{scope}
\begin{scope}
\path[clip] (  0.00,  0.00) rectangle (216.81,144.54);
\definecolor{drawColor}{gray}{0.20}

\path[draw=drawColor,line width= 0.6pt,line join=round] ( 38.75, 27.94) --
	( 38.75, 30.69);

\path[draw=drawColor,line width= 0.6pt,line join=round] ( 50.49, 27.94) --
	( 50.49, 30.69);

\path[draw=drawColor,line width= 0.6pt,line join=round] ( 62.22, 27.94) --
	( 62.22, 30.69);

\path[draw=drawColor,line width= 0.6pt,line join=round] ( 73.95, 27.94) --
	( 73.95, 30.69);

\path[draw=drawColor,line width= 0.6pt,line join=round] ( 85.69, 27.94) --
	( 85.69, 30.69);

\path[draw=drawColor,line width= 0.6pt,line join=round] ( 97.42, 27.94) --
	( 97.42, 30.69);

\path[draw=drawColor,line width= 0.6pt,line join=round] (109.15, 27.94) --
	(109.15, 30.69);

\path[draw=drawColor,line width= 0.6pt,line join=round] (120.89, 27.94) --
	(120.89, 30.69);

\path[draw=drawColor,line width= 0.6pt,line join=round] (132.62, 27.94) --
	(132.62, 30.69);

\path[draw=drawColor,line width= 0.6pt,line join=round] (144.35, 27.94) --
	(144.35, 30.69);
\end{scope}
\begin{scope}
\path[clip] (  0.00,  0.00) rectangle (216.81,144.54);
\definecolor{drawColor}{gray}{0.30}

\node[text=drawColor,anchor=base,inner sep=0pt, outer sep=0pt, scale=  0.88] at ( 38.75, 19.68) {0};

\node[text=drawColor,anchor=base,inner sep=0pt, outer sep=0pt, scale=  0.88] at ( 50.49, 19.68) {10};

\node[text=drawColor,anchor=base,inner sep=0pt, outer sep=0pt, scale=  0.88] at ( 62.22, 19.68) {20};

\node[text=drawColor,anchor=base,inner sep=0pt, outer sep=0pt, scale=  0.88] at ( 73.95, 19.68) {30};

\node[text=drawColor,anchor=base,inner sep=0pt, outer sep=0pt, scale=  0.88] at ( 85.69, 19.68) {40};

\node[text=drawColor,anchor=base,inner sep=0pt, outer sep=0pt, scale=  0.88] at ( 97.42, 19.68) {50};

\node[text=drawColor,anchor=base,inner sep=0pt, outer sep=0pt, scale=  0.88] at (109.15, 19.68) {60};

\node[text=drawColor,anchor=base,inner sep=0pt, outer sep=0pt, scale=  0.88] at (120.89, 19.68) {70};

\node[text=drawColor,anchor=base,inner sep=0pt, outer sep=0pt, scale=  0.88] at (132.62, 19.68) {80};

\node[text=drawColor,anchor=base,inner sep=0pt, outer sep=0pt, scale=  0.88] at (144.35, 19.68) {90};
\end{scope}
\begin{scope}
\path[clip] (  0.00,  0.00) rectangle (216.81,144.54);
\definecolor{drawColor}{RGB}{0,0,0}

\node[text=drawColor,anchor=base,inner sep=0pt, outer sep=0pt, scale=  1.10] at ( 91.55,  7.64) {WER, \%};
\end{scope}
\begin{scope}
\path[clip] (  0.00,  0.00) rectangle (216.81,144.54);
\definecolor{drawColor}{RGB}{0,0,0}

\node[text=drawColor,rotate= 90.00,anchor=base,inner sep=0pt, outer sep=0pt, scale=  1.10] at ( 13.08, 84.86) {Accuracy, \%};
\end{scope}
\begin{scope}
\path[clip] (  0.00,  0.00) rectangle (216.81,144.54);
\definecolor{fillColor}{RGB}{255,255,255}

\path[fill=fillColor] (162.39, 57.30) rectangle (211.31,112.42);
\end{scope}
\end{tikzpicture}
	\vspace*{-0.2in}
	\caption{MRDA}
	\end{subfigure}
	\begin{subfigure}{.3\textwidth}
	\centering
\begin{tikzpicture}[x=1pt,y=1pt]
\definecolor{fillColor}{RGB}{255,255,255}
\path[use as bounding box,fill=fillColor,fill opacity=0.00] (0,0) rectangle (216.81,144.54);
\begin{scope}
\path[clip] (  0.00,  0.00) rectangle (216.81,144.54);
\definecolor{drawColor}{RGB}{255,255,255}
\definecolor{fillColor}{RGB}{255,255,255}

\path[draw=drawColor,line width= 0.6pt,line join=round,line cap=round,fill=fillColor] (  0.00,  0.00) rectangle (216.81,144.54);
\end{scope}
\begin{scope}
\path[clip] ( 36.11, 30.69) rectangle (151.39,139.04);
\definecolor{fillColor}{gray}{0.92}

\path[fill=fillColor] ( 36.11, 30.69) rectangle (151.39,139.04);
\definecolor{drawColor}{RGB}{255,255,255}

\path[draw=drawColor,line width= 0.3pt,line join=round] ( 36.11, 47.92) --
	(151.39, 47.92);

\path[draw=drawColor,line width= 0.3pt,line join=round] ( 36.11, 72.55) --
	(151.39, 72.55);

\path[draw=drawColor,line width= 0.3pt,line join=round] ( 36.11, 97.18) --
	(151.39, 97.18);

\path[draw=drawColor,line width= 0.3pt,line join=round] ( 36.11,121.80) --
	(151.39,121.80);

\path[draw=drawColor,line width= 0.6pt,line join=round] ( 36.11, 35.61) --
	(151.39, 35.61);

\path[draw=drawColor,line width= 0.6pt,line join=round] ( 36.11, 60.24) --
	(151.39, 60.24);

\path[draw=drawColor,line width= 0.6pt,line join=round] ( 36.11, 84.86) --
	(151.39, 84.86);

\path[draw=drawColor,line width= 0.6pt,line join=round] ( 36.11,109.49) --
	(151.39,109.49);

\path[draw=drawColor,line width= 0.6pt,line join=round] ( 36.11,134.11) --
	(151.39,134.11);

\path[draw=drawColor,line width= 0.6pt,line join=round] ( 44.55, 30.69) --
	( 44.55,139.04);

\path[draw=drawColor,line width= 0.6pt,line join=round] ( 58.61, 30.69) --
	( 58.61,139.04);

\path[draw=drawColor,line width= 0.6pt,line join=round] ( 72.66, 30.69) --
	( 72.66,139.04);

\path[draw=drawColor,line width= 0.6pt,line join=round] ( 86.72, 30.69) --
	( 86.72,139.04);

\path[draw=drawColor,line width= 0.6pt,line join=round] (100.78, 30.69) --
	(100.78,139.04);

\path[draw=drawColor,line width= 0.6pt,line join=round] (114.84, 30.69) --
	(114.84,139.04);

\path[draw=drawColor,line width= 0.6pt,line join=round] (128.90, 30.69) --
	(128.90,139.04);

\path[draw=drawColor,line width= 0.6pt,line join=round] (142.96, 30.69) --
	(142.96,139.04);
\definecolor{fillColor}{RGB}{68,1,84}

\path[fill=fillColor] ( 38.22, 35.61) rectangle ( 44.55,134.11);
\definecolor{fillColor}{RGB}{253,231,37}

\path[fill=fillColor] ( 44.55, 35.61) rectangle ( 50.87,128.84);
\definecolor{fillColor}{RGB}{68,1,84}

\path[fill=fillColor] ( 52.28, 35.61) rectangle ( 58.61,129.55);
\definecolor{fillColor}{RGB}{253,231,37}

\path[fill=fillColor] ( 58.61, 35.61) rectangle ( 64.93,125.91);
\definecolor{fillColor}{RGB}{68,1,84}

\path[fill=fillColor] ( 66.34, 35.61) rectangle ( 72.66,109.09);
\definecolor{fillColor}{RGB}{253,231,37}

\path[fill=fillColor] ( 72.66, 35.61) rectangle ( 78.99,116.76);
\definecolor{fillColor}{RGB}{68,1,84}

\path[fill=fillColor] ( 80.40, 35.61) rectangle ( 86.72,106.56);
\definecolor{fillColor}{RGB}{253,231,37}

\path[fill=fillColor] ( 86.72, 35.61) rectangle ( 93.05,109.32);
\definecolor{fillColor}{RGB}{68,1,84}

\path[fill=fillColor] ( 94.46, 35.61) rectangle (100.78,108.86);
\definecolor{fillColor}{RGB}{253,231,37}

\path[fill=fillColor] (100.78, 35.61) rectangle (107.11,111.38);
\definecolor{fillColor}{RGB}{68,1,84}

\path[fill=fillColor] (108.51, 35.61) rectangle (114.84, 98.70);
\definecolor{fillColor}{RGB}{253,231,37}

\path[fill=fillColor] (114.84, 35.61) rectangle (121.17,103.12);
\definecolor{fillColor}{RGB}{68,1,84}

\path[fill=fillColor] (122.57, 35.61) rectangle (128.90,113.94);
\definecolor{fillColor}{RGB}{253,231,37}

\path[fill=fillColor] (128.90, 35.61) rectangle (135.23,116.31);
\definecolor{fillColor}{RGB}{68,1,84}

\path[fill=fillColor] (136.63, 35.61) rectangle (142.96,109.49);
\definecolor{fillColor}{RGB}{253,231,37}

\path[fill=fillColor] (142.96, 35.61) rectangle (149.28,109.49);
\end{scope}
\begin{scope}
\path[clip] (  0.00,  0.00) rectangle (216.81,144.54);
\definecolor{drawColor}{gray}{0.30}

\node[text=drawColor,anchor=base east,inner sep=0pt, outer sep=0pt, scale=  0.88] at ( 31.16, 32.58) {0};

\node[text=drawColor,anchor=base east,inner sep=0pt, outer sep=0pt, scale=  0.88] at ( 31.16, 57.21) {25};

\node[text=drawColor,anchor=base east,inner sep=0pt, outer sep=0pt, scale=  0.88] at ( 31.16, 81.83) {50};

\node[text=drawColor,anchor=base east,inner sep=0pt, outer sep=0pt, scale=  0.88] at ( 31.16,106.46) {75};

\node[text=drawColor,anchor=base east,inner sep=0pt, outer sep=0pt, scale=  0.88] at ( 31.16,131.08) {100};
\end{scope}
\begin{scope}
\path[clip] (  0.00,  0.00) rectangle (216.81,144.54);
\definecolor{drawColor}{gray}{0.20}

\path[draw=drawColor,line width= 0.6pt,line join=round] ( 33.36, 35.61) --
	( 36.11, 35.61);

\path[draw=drawColor,line width= 0.6pt,line join=round] ( 33.36, 60.24) --
	( 36.11, 60.24);

\path[draw=drawColor,line width= 0.6pt,line join=round] ( 33.36, 84.86) --
	( 36.11, 84.86);

\path[draw=drawColor,line width= 0.6pt,line join=round] ( 33.36,109.49) --
	( 36.11,109.49);

\path[draw=drawColor,line width= 0.6pt,line join=round] ( 33.36,134.11) --
	( 36.11,134.11);
\end{scope}
\begin{scope}
\path[clip] (  0.00,  0.00) rectangle (216.81,144.54);
\definecolor{drawColor}{gray}{0.20}

\path[draw=drawColor,line width= 0.6pt,line join=round] ( 44.55, 27.94) --
	( 44.55, 30.69);

\path[draw=drawColor,line width= 0.6pt,line join=round] ( 58.61, 27.94) --
	( 58.61, 30.69);

\path[draw=drawColor,line width= 0.6pt,line join=round] ( 72.66, 27.94) --
	( 72.66, 30.69);

\path[draw=drawColor,line width= 0.6pt,line join=round] ( 86.72, 27.94) --
	( 86.72, 30.69);

\path[draw=drawColor,line width= 0.6pt,line join=round] (100.78, 27.94) --
	(100.78, 30.69);

\path[draw=drawColor,line width= 0.6pt,line join=round] (114.84, 27.94) --
	(114.84, 30.69);

\path[draw=drawColor,line width= 0.6pt,line join=round] (128.90, 27.94) --
	(128.90, 30.69);

\path[draw=drawColor,line width= 0.6pt,line join=round] (142.96, 27.94) --
	(142.96, 30.69);
\end{scope}
\begin{scope}
\path[clip] (  0.00,  0.00) rectangle (216.81,144.54);
\definecolor{drawColor}{gray}{0.30}

\node[text=drawColor,anchor=base,inner sep=0pt, outer sep=0pt, scale=  0.88] at ( 44.55, 19.68) {0};

\node[text=drawColor,anchor=base,inner sep=0pt, outer sep=0pt, scale=  0.88] at ( 58.61, 19.68) {10};

\node[text=drawColor,anchor=base,inner sep=0pt, outer sep=0pt, scale=  0.88] at ( 72.66, 19.68) {20};

\node[text=drawColor,anchor=base,inner sep=0pt, outer sep=0pt, scale=  0.88] at ( 86.72, 19.68) {30};

\node[text=drawColor,anchor=base,inner sep=0pt, outer sep=0pt, scale=  0.88] at (100.78, 19.68) {40};

\node[text=drawColor,anchor=base,inner sep=0pt, outer sep=0pt, scale=  0.88] at (114.84, 19.68) {50};

\node[text=drawColor,anchor=base,inner sep=0pt, outer sep=0pt, scale=  0.88] at (128.90, 19.68) {60};

\node[text=drawColor,anchor=base,inner sep=0pt, outer sep=0pt, scale=  0.88] at (142.96, 19.68) {70};
\end{scope}
\begin{scope}
\path[clip] (  0.00,  0.00) rectangle (216.81,144.54);
\definecolor{drawColor}{RGB}{0,0,0}

\node[text=drawColor,anchor=base,inner sep=0pt, outer sep=0pt, scale=  1.10] at ( 93.75,  7.64) {WER, \%};
\end{scope}
\begin{scope}
\path[clip] (  0.00,  0.00) rectangle (216.81,144.54);
\definecolor{drawColor}{RGB}{0,0,0}

\node[text=drawColor,rotate= 90.00,anchor=base,inner sep=0pt, outer sep=0pt, scale=  1.10] at ( 13.08, 84.86) {Accuracy, \%};
\end{scope}
\begin{scope}
\path[clip] (  0.00,  0.00) rectangle (216.81,144.54);
\definecolor{fillColor}{RGB}{255,255,255}

\path[fill=fillColor] (162.39, 57.30) rectangle (211.31,112.42);
\end{scope}
\begin{scope}
\path[clip] (  0.00,  0.00) rectangle (216.81,144.54);
\definecolor{drawColor}{RGB}{0,0,0}

\node[text=drawColor,anchor=base west,inner sep=0pt, outer sep=0pt, scale=  1.10] at (167.89, 98.28) {Architecture};
\end{scope}
\begin{scope}
\path[clip] (  0.00,  0.00) rectangle (216.81,144.54);
\definecolor{fillColor}{gray}{0.95}

\path[fill=fillColor] (167.89, 77.26) rectangle (182.35, 91.71);
\end{scope}
\begin{scope}
\path[clip] (  0.00,  0.00) rectangle (216.81,144.54);
\definecolor{fillColor}{RGB}{68,1,84}

\path[fill=fillColor] (168.60, 77.97) rectangle (181.64, 91.00);
\end{scope}
\begin{scope}
\path[clip] (  0.00,  0.00) rectangle (216.81,144.54);
\definecolor{fillColor}{gray}{0.95}

\path[fill=fillColor] (167.89, 62.80) rectangle (182.35, 77.26);
\end{scope}
\begin{scope}
\path[clip] (  0.00,  0.00) rectangle (216.81,144.54);
\definecolor{fillColor}{RGB}{253,231,37}

\path[fill=fillColor] (168.60, 63.51) rectangle (181.64, 76.54);
\end{scope}
\begin{scope}
\path[clip] (  0.00,  0.00) rectangle (216.81,144.54);
\definecolor{drawColor}{RGB}{0,0,0}

\node[text=drawColor,anchor=base west,inner sep=0pt, outer sep=0pt, scale=  0.88] at (187.85, 81.45) {Pipeline};
\end{scope}
\begin{scope}
\path[clip] (  0.00,  0.00) rectangle (216.81,144.54);
\definecolor{drawColor}{RGB}{0,0,0}

\node[text=drawColor,anchor=base west,inner sep=0pt, outer sep=0pt, scale=  0.88] at (187.85, 67.00) {End-to-end};
\end{scope}
\end{tikzpicture}
	\vspace*{-0.2in}
	\caption{FSC}
	\end{subfigure}
	\caption{Accuracy comparison for the utterances grouped by ASR WER}
\label{fig:acc}
\end{figure*}

\subsubsection{Objective function}

Comparison of objective functions on downstream tasks,
as well as cross-comparison of how selected objective function
influences value of others on validation subset,
is provided in \mbox{Table}~\ref{tab:loss}.
Overall, these results indicate that the evaluated
objective functions behave similarly in this task,
however L1 distance based objective function
yields slightly better results.

\begin{table}[h]
  \caption{Effect of objective function and longer training}
  \label{tab:loss}
  \centering
  \footnotesize
  \begin{tabular*}{1.01\columnwidth}{l|c|c|c|c|c|c}
    \noalign{\hrule height 1pt}
	  Objective & \multicolumn{3}{|c|}{Accuracy on Test, \%} & \multicolumn{3}{|c}{Validation value} \\
    \cline{2-7}
	  function & SwBD & MRDA & FSC & Cosine & L2 & L1 \\
    \hline
	  Cosine & 55.56 & 59.64 & 89.45 & 0.13 & 0.08 & 0.21 \\
	  L2 & 53.73 & 59.91 & 88.64 & 0.13 & 0.07 & 0.20 \\
	  L1 & 56.32 & \textbf{60.39} & 89.98 & 0.13 & 0.07 & 0.20 \\
    \hline
	  L1, 61 ep. & \textbf{58.60} & 60.18 & \textbf{91.12} & 0.11 & 0.06 & 0.18 \\
    \noalign{\hrule height 1pt}
  \end{tabular*}
\end{table}

\subsection{Further supervised fine-tuning on downstream tasks}

The resulting end-to-end utterance classification model
is fully differentiable and can be further
optimized for a downstream task
by applying standard supervised neural network training methods
with few labeled speech samples.
This feature should be helpful for the full exploitation
of information that is relevant for
the task and is encoded in speech,
what is less trivial to implement in the traditional
ASR and NLU pipeline setup, where intermediate representation
has to be discrete and for example flatten
the rich variety of prosodic events
to few punctuation characters.
We examine whether it is useful in practice
by running standard supervised classifier
training on few samples from training subsets.
Table \ref{tab:downstream}
compares the results of fine-tuning of the output layer
alone and together with two former ASR encoder layers.
We conclude that end-to-end approach indeed can
overcome the error propagation problem of pipeline SLU approach
by the automatic propagation of the error signal back to
relevant parts of SLU system. However,
additional training samples may sometimes easily skew
the small training dataset away from the testing dataset
and cause worse results, so more attention should be payed
to the selection of training samples.

\begin{table}[h]
	\caption{
	Effect of supervised fine-tuning on downstream tasks
	}
  \label{tab:downstream}
  \centering
  \footnotesize
  \begin{tabular*}{1.045\columnwidth}{l|c|c|c|c|c|c}
    \noalign{\hrule height 1pt}
	  Num. of & \multicolumn{6}{|c}{Fine-tuned layers (accuracy on Test, \%)} \\
    \cline{2-7}
	  samples & \multicolumn{3}{|c|}{Output layer} & \multicolumn{3}{|c}{Output and hidden layers}  \\
    \cline{2-7}
	  per class & SwBD & MRDA & FSC & SwBD & MRDA & FSC \\
    \hline
	  0 & 58.60 & 60.18 & 91.12 & 58.60 & 60.18 & 91.12 \\
	\hline
	  1 & 58.60 & 60.59 & 93.62 & 58.60 & 60.41 & 94.15 \\
	  2 & 58.60 & 60.22 & 93.44 & 58.60 & 60.40 & 95.04 \\
	  3 & 58.83 & 60.22 & 93.33 & 58.83 & 60.16 & 94.83 \\
	  4 & 58.55 & 60.35 & \textbf{93.96} & 58.71 & 60.47 & \textbf{95.54} \\
	  10 & \textbf{60.14} & \textbf{60.94} & 93.88 & \textbf{60.22} & \textbf{61.32} & 95.49 \\
    \hline
    \noalign{\hrule height 1pt}
  \end{tabular*}
\end{table}

\section{Qualitative Analysis}

We attempt to assess the differences between
the pipeline and end-to-end SLU approaches in greater detail
by looking at the accuracy values on groups of utterances
split by the WER levels of the baseline ASR system.
Our hypothesis is that the pipeline system
would make more mistakes on the more challenging recordings
characterized by higher WER values, because
ASR systems are optimized for phonetic or graphemic
similarity to the ground truth and are more likely to lose
semantic information in case of errors.
Figure \ref{fig:acc} shows the accuracy values of the 
pipeline system and the best end-to-end SLU model
(without fine-tuning).
The utterances are grouped by WER of the baseline ASR
in 10\% ranges. The ranges with WER $>$ 100\%
are not included for brevity, they account for
487, 3847 and 44 utterances in SwBD, MRDA and FSC testing
subsets respectively.
The results confirm fully our hypothesis on FSC dataset
and to some extent on other two datasets.

\section{Conclusions}
We proposed to combine parameters transfer
from well trained ASR and NLU models with
Teacher-Student learning for final alignment
of SLU output space to NLU output space
in order to construct end-to-end SLU model
allowing few-shot transfer of downstream tasks
from text to speech.
We outlined necessary steps and settings for
the practical pretrained NLU model adaptation in SLU via cross-modal transfer.
Our system reaches accuracy of
58.60\%, 60.18\% and 91.12\%
on SwBD, MRDA and FSC datasets
without fine-tuning
and 60.22\%, 61.32\% and 95.49\%
after fine-tuning on ten labeled samples per class
compared to 57.23\%, 64.06\% and 94.57\% reached by
the pipeline system.
The results of this research support the idea that text pretrained
contextual embeddings can be useful for tasks outside of text modality.
The present study also adds new tasks to the growing body of
research of language processing methods using Transformer neural networks.

\bibliographystyle{IEEEtran}
\bibliography{paper}

\end{document}